%% file: 00ijcai26.tex
\documentclass{article}
\usepackage{ijcai26}

\usepackage{times}
\usepackage{soul}
\usepackage{url}
\usepackage[hidelinks]{hyperref}
\usepackage[utf8]{inputenc}
\usepackage[small]{caption}
\usepackage{graphicx}
\usepackage{amsmath}
\usepackage{amsthm}
\usepackage{booktabs}
\usepackage{algorithm}
\usepackage{algorithmic}
\usepackage[switch]{lineno}
\usepackage{enumitem}
\usepackage{makecell}
\usepackage{xcolor}
\usepackage{multirow}
\usepackage{subcaption}
\usepackage{algorithmic}
\usepackage{amssymb, amsfonts}
\usepackage{placeins}
\usepackage[normalem]{ulem}

\title{Momentum-integrated Multi-task Stock Recommendation with \\ Converge-based Optimization}

\author{
Hao Wang$^1$\and
Jingshu Peng$^2$\and
Yanyan Shen$^{3}$\and
Xujia Li$^{1,2}$\and
Quanqing Xu$^4$\and
Chuanhui Yang$^4$\And
Lei Chen$^{1,2}$
\\
\affiliations
$^1$HKUST(GZ),
$^2$HKUST,
$^3$Shanghai Jiao Tong University,
$^4$OceanBase, Ant Group\\
\emails
seraveea@connect.hkust-gz.edu.cn,
jpengab@cse.ust.hk,
\{leexujia, leichen\}@ust.hk,
shenyy@sjtu.edu.cn,
\{xuquanqing.xqq, rizhao.ych\}@oceanbase.com,
}

\begin{document}

\maketitle

\begin{abstract}
Stock recommendation is critical in Fintech applications, which leverage price series and alternative information to estimate future stock performance. 
Traditional time-series forecasting training often fails to capture stock trends and rankings simultaneously, which are essential factors for investors. To tackle this issue, we introduce a Multi-Task Learning (MTL) framework for stock recommendation, \textbf{M}omentum-\textbf{i}ntegrated \textbf{M}ulti-task \textbf{Stoc}k \textbf{R}ecommendation with Converge-based Optimization (\textbf{MiM-StocR}). To improve the model's ability to capture short-term trends, we incorporate a momentum line indicator in model training. To prioritize top-performing stocks and optimize investment allocation, we propose a listwise ranking loss function called Adaptive-k ApproxNDCG. Moreover, due to the volatility and uncertainty of the stock market, existing MTL frameworks face overfitting issues when applied to stock time series. To mitigate this issue, we introduce the Converge-based Quad-Balancing (CQB) method. We conducted extensive experiments on three stock benchmarks: SEE50, CSI 100, and CSI 300. MiM-StocR outperforms state-of-the-art MTL baselines across both ranking and profitability evaluations.
\end{abstract}

\section{Introduction}
\label{sec:intro}
\input{01new_intro.tex}

\section{Related Work}
\input{05related_work.tex}

\section{Problem Statement}
\input{02ps.tex}

\section{Methodology}
\input{03method.tex}

\section{Experiments}
\input{04experiment.tex}

\section{Conclusions}
\input{06conclusion.tex}

\bibliographystyle{named}
\bibliography{ijcai26}


\end{document}

%% file: 01new_intro.tex
Deep learning has become a prominent approach for stock recommendation at the intersection of artificial intelligence and finance~\cite{rather2015recurrent,xu2021hist,wang2023methods}.
Quantitative investors increasingly rely on deep models to predict price-related indicators and design investment strategies.
However, despite their empirical success, existing deep learning methods remain misaligned with real-world quantitative investment practices in terms of problem formulation, training objectives.
This mismatch poses bottlenecks for deploying deep models in practical investment.

\begin{figure}
    \centering
    \includegraphics[width=0.7\linewidth]{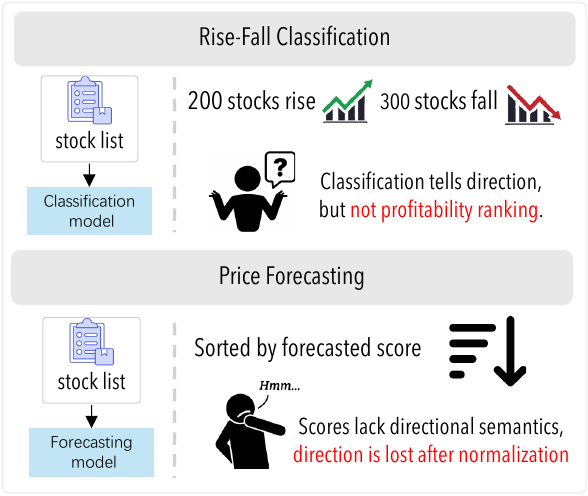}
    \caption{Neither classification nor regression provides actionable guidance for profit-oriented stock selection.}
    \label{fig:intro}
\end{figure}

First, conventional objectives provide limited insights for investors, who care not only about price movements of stocks but also about their relative rankings and profit potential. As shown in Figure~\ref{fig:intro},
most existing works formulate stock recommendation as classification (e.g., rise-fall prediction) or regression (e.g., price forecasting) tasks.
Due to the high volatility and noise in markets~\cite{lee2012jumps}, rise-fall prediction performance often remains close to random guessing~\cite{hu2018listening}, while regression outputs are typically regularized and lack direction for investment decisions~\cite{zou2022stock,rather2015recurrent}.
Neither task alone produces sufficiently informative signals for practical portfolio construction.

Second, real-world quantitative investment prioritizes ranking stocks to allocate limited capital toward the most promising candidates.
Classic factor-based models, such as the Fama--French framework~\cite{fama1993common}, explicitly rely on ranking stocks by financial factors.
Although recent studies introduce ranking losses~\cite{feng2019temporal,sawhney2021stock}, these methods treat all stock pairs equally and dilute the importance of top-ranked stocks, which are the primary focus in portfolio design~\cite{heinrich2021factor,Becker2018State}.

Third, stock time series exhibit severe distribution shifts between training and future data~\cite{bhowmik2020stock,zhao2023doubleadapt}, making deep models highly prone to overfitting.
As illustrated in Figure~\ref{fig:overfit}-A, validation and test performance often stagnate after early epochs, highlighting the necessity of explicitly addressing overfitting during training.

To address these challenges, we develop MiM-StocR, a multi-task stock forecasting framework designed to better align model training with practical investment requirements.
The framework adopts a multi-task learning paradigm that jointly considers regression and classification objectives, and incorporates a momentum-line indicator as a trend-aware signal to alleviate the noise inherent in binary rise-or-fall labels.
Momentum is a well-established investment factor~\cite{jegadeesh1993returns,asness2014fact} that captures persistent price trends and has been shown to be effective over long horizons.
We discretize the momentum line into five levels and formulate it as a multi-class classification task, which provides a more structured representation and facilitates more stable learning for ranking.

To better reflect ranking-oriented investment decisions, we adopt an adaptive list-wise ranking objective, Adaptive-$k$ ApproxNDCG, which emphasizes the relative ordering of top-tier stocks.
NDCG enables direct optimization of ranking quality at the top of the list.
We incorporate an adaptive-$k$ strategy that adjusts the truncation boundary according to the distribution of momentum levels, leading to more consistent ranking supervision and improved robustness.

Finally, to handle overfitting under distribution shift and imbalance in multi-task optimization~\cite{yu2020gradient,kim2021reversible},
we implement a convergence-aware optimization strategy, referred to as Converge-based Quad-Balancing (CQB).
CQB regulates gradient smoothing and regularization strength based on the relative convergence behavior between training and split meta-validation losses, with the goal of stabilizing training dynamics and reducing the influence of overfitted gradients in multi-task settings.

We evaluate MiM-StocR on three real-world benchmarks, SEE50, CSI100, and CSI300.
Extensive experiments and ablation studies demonstrate consistent improvements in ranking performance and profitability.
Our code is publicly available at an anonymous repository.\footnote{\url{https://anonymous.4open.science/r/MiM-StocR-E6BA}}

%% file: 05related_work.tex
\label{sec:related_work}

\textbf{Stock Recommendation.}
Stock recommendation has been widely studied, ranging from classical time-series models to deep neural networks~\cite{ariyo2014stock,rather2015recurrent,bao2017deep,ding2020hierarchical}. 
Recent work further incorporates alternative data sources, such as financial news, knowledge graphs, and relational structures, to improve performance~\cite{ding2015deep,cheng2020knowledge,wang2023methods,ong2024deepunifiedmom}. 
Graph-based methods, such as RSR~\cite{feng2019temporal} and HIST~\cite{xu2021hist}, further exploit inter-stock relations and remain strong baselines in practice~\cite{yang2020qlib}.
\cite{alzaman2025optimizing} implements reinforcement learning on stock recommendation. In this work, we focus on supervised deep learning scenario of stock recommendation.

\textbf{Noisy Trend Prediction and Momentum.}
Binary rise-or-fall prediction is known to be noisy and difficult to scale to large stock universes~\cite{hu2018listening,long2020integrated}. 
Momentum-based representations provide a more stable supervisory signal and have demonstrated consistent effectiveness in both empirical finance and learning-based models~\cite{asness2014fact}. 
Recent studies further integrate momentum modeling into multi-task learning frameworks for portfolio construction, such as DeepUnifiedMom~\cite{ong2024deepunifiedmom}.

\textbf{Rank-aware Objectives in Stock Recommendation.}
Rank-aware learning has been introduced to better align optimization objectives with investment returns. 
Existing approaches mainly rely on pairwise or listwise ranking losses in both financial and general multi-task learning settings~\cite{feng2019temporal,sawhney2021stock,kwiatkowski2025evaluating}. 
\cite{wu2024momentum} uses learning-to-rank for momentum-based portfolio selection with heterogeneous knowledge graphs.

\textbf{Multi-task Learning and Optimization.}
Multi-task learning jointly optimizes multiple objectives but often suffers from gradient conflict and imbalance~\cite{sener2018multi,yu2020gradient,liu2021conflict,navon2022multi,durmus2024pairwise}. 
In financial applications, MTL is commonly used to jointly model price regression, trend prediction or risk-related tasks~\cite{yang2020html,ma2022stock,ma2024quantitative,ong2023constructing}.
\cite{he2022metabalance} addresses gradient magnitude imbalance in multi-task recommendation systems via adaptive balancing.
Related work has explored multi-trend objectives from an optimization perspective using Quasi-Newton methods~\cite{lin2024portfolio}.
\cite{ma2024quantitative} jointly model stock return and volatility risk via multi-task learning for portfolio optimization.
However, existing MTL optimizers often struggle under distribution shift and overfitting in financial time series.

%% file: 02ps.tex
We first define the stock recommendation problem.
Following by previous work~\cite{hu2018listening}, we define the stock recommend score as the one-day return ratio:

\begin{equation}
    y_{i}^t = \frac{price_{i}^{t+1}-price_{i}^{t}}{price_{i}^{t}}
    \label{eq:definition_score}
\end{equation}
where $price_{i}^{t}$ is the closing price of stock $i$ in trading day $t$.
Given the feature set of all stocks on date $t$, the objective of the stock recommendation is to predict the one-day return ratio for each stock and rank them based on the predicted values, aiming to produce a ranking that closely aligns with the true order and achieve better profits in investment.




%% file: 03method.tex
\begin{figure*}
    \centering
    \includegraphics[width=0.9\linewidth]{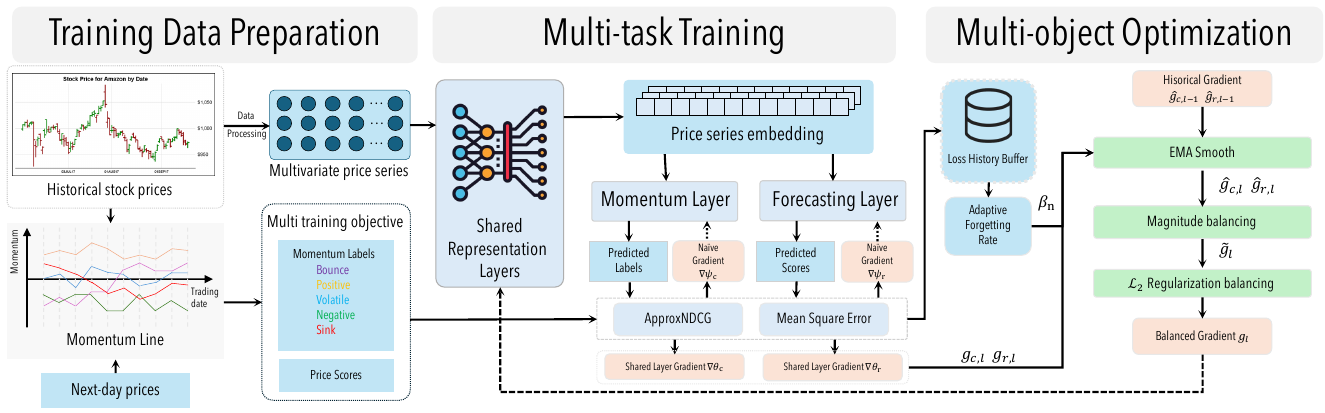}
    \caption{Overview of the MiM-StocR framework. It integrates momentum-based multi-task learning with Adaptive-k ApproxNDCG ranking loss and CQB optimization for robust stock recommendation.}
    \vspace{-0.4cm}
    \label{fig:demo}
\end{figure*}

We introduce the workflow of MiM-StocR in Figure~\ref{fig:demo}, the framework has three stages: Data Preparation, Multi-Task Learning, and Multi-Object Optimization.

In the Task Preparation stage, raw time series are processed, and the training objective is generated for each task to align with real-world investment scenarios.
The first task is forecasting the one-day return ratio described in the Problem Statement.
For the second task, to improve the model’s perception of price trends, we propose a momentum line indicator as the ground truth for the second sub-task.

In the Multi-Task Learning stage, the two tasks are trained under a hard parameter-sharing MTL structure~\cite{zhang2021survey}. 
To introduce the ranking and emphasize top-tier stocks, we implement the Adaptive-k ApproxNDCG as the objective function for classification.

In the Multi-Objective Optimization stage, we introduce Converge-based Quad Balancing (CQB), which can mitigate the overfitting issue and balance magnitudes of different losses and gradients in Section~\ref{sec:cqb}.

\subsection{Momentum Line Indicator Construction}
\label{sec:mtm}
Given that the financial market is fluctuating, there are doubts about whether learning to predict rise-or-fall could benefit the stock recommendation model~\cite{bengio1997using}.
Therefore, we adopt the momentum line indicator and formulate a task to replace the noisy rise-or-fall signal.
Momentum investing is a strategy that aims to capitalize on the continuance of existing market trends.
The effectiveness of momentum investing has been proven and widely applied in real stock market~\cite{asness2014fact,barroso2015momentum,jegadeesh1993returns}.
We compute the price momentum formula by the following equation:
\begin{equation}
    m_T = {price}_T - {price}_{T-l}
    \label{eq:momentum formula}
\end{equation}
$m_T$ is the momentum at day $T$, $price_T$ is the closing price at day $T$ and $l$ is the length of gap days in momentum. To represent the short-term trend, we build a momentum line, which is formed by a list of momentum values: $\{m_{T-s}, m_{T-s-1}, ..., m_{T}\}$,
and $s$ is the length of the momentum line.
Based on the trend of the momentum line, we categorize all momentum lines into five levels that serve as the label in the classification task:
\begin{itemize}
\item \textbf{``Bounce"} The line changes from negative to positive. 
\item \textbf{``Positive"} The line stays positive. 
\item \textbf{``Volatile"} The line oscillates around 0.
\item \textbf{``Negative"} The line stays negative. 
\item \textbf{``Sink"} The line changes from positive to negative. 
\end{itemize}


The shapes of momentum line are shown in Figure~\ref{fig:demo}. 
High momentum levels indicate that stocks have a strong positive momentum in a short period.
From the perspective of economics, momentum investing is a profitable short-term investing strategy. 
Since momentum investment strategies are positively related to higher profits in real market trading, we could suppose that the momentum is related to return ratio. 

\subsection{Adaptive-k ApproxNDCG}
\label{sec:ndcg}

To incorporate ranking information into training and emphasize top-performing stocks, we propose a NDCG@k-based objective, termed Adaptive-k ApproxNDCG.
Normalized Discounted Cumulative Gain (NDCG) is defined as
\begin{equation}
    NDCG(\pi_f, w)=\frac{DCG(\pi_f,w)}{DCG(\pi_f^{*},w)},
    \label{eq:ndcg_0}
\end{equation}
where $\pi_f$ denotes the ranking induced by the scoring function $f$, $w$ represents item relevance weights, and $\pi_f^{*}$ is the ideal ranking.
The discounted cumulative gain (DCG) is given by
\begin{equation}
    DCG(\pi_f, w)=\sum_{i=1}^{n}\frac{2^{w^i-1}}{\log_2(1+\pi(i))},
    \label{eq:ndcg_1}
\end{equation}
with the rank position:
\begin{equation}
    \pi(i) \triangleq 1 + \sum_{j \ne i}\mathbb{I}_{f(i) < f(j)},
    \label{eq:rank_function}
\end{equation}
where $\mathbb{I}_{f(i)<f(j)}$ is a binary indicator.
Since this indicator is non-differentiable, NDCG cannot be directly optimized.

Following~\cite{qin2010general}, we replace the indicator with a smooth sigmoid approximation:
\begin{equation}
    \mathbb{I}_{f(i)<f(j)} \triangleq \frac{1}{1+e^{f(i)-f(j)}},
    \label{eq:sigomoid indicator}
\end{equation}
which enables gradient-based optimization.
To focus on top-tier stocks, we optimize ApproxNDCG@$k$.
However, the number of high-momentum stocks varies across trading days, and using a fixed $k$ may exclude stocks with identical momentum levels, leading to truncation effects~\cite{wang2013theoretical}.
To address this issue, we adaptively determine $k$ using a lower-bound threshold $\tau$:
\begin{equation}
    k = \sum_{j=4} |G_{j}| ,
    \label{eq:adaptive-k}
\end{equation}
where $k$ is the minimal integer satisfying $k \geq \tau$, and $|G_j|$ denotes the number of stocks in momentum level $j$, starting from the highest level defined in Section~\ref{sec:mtm}.
Stock groups are iteratively included until the threshold is met, ensuring that stocks within the same momentum level are not split by the truncation.
The resulting loss is defined as
\begin{equation}
    \mathcal{L}_{ndcg} = e^{-ApproxNDCG(\pi_{w_{pred}}, w, k)},
\end{equation}
where $\pi_{w_{pred}}$ is the ranking induced by predicted weights and $k$ is adaptively computed via Eq.~\ref{eq:adaptive-k}.
This exponential form preserves differentiability and monotonicity, assigning lower loss to rankings closer to the label-induced ideal ranking.

Finally, to jointly optimize classification accuracy and ranking quality, we combine cross-entropy loss with Adaptive-k ApproxNDCG:
\begin{equation}
    \mathcal{L}_{c} = \lambda_{ce}\mathcal{L}_{ce} + (1-\lambda_{ce})\mathcal{L}_{ndcg},
\end{equation}
where $\lambda_{ce} \in [0,1]$ controls the trade-off.
We set $\lambda_{ce}=0.5$ in all experiments, which yields stable convergence and robust performance across datasets by dynamically aligning the optimization focus with daily market conditions.



\subsection{Converge-based Quad-Balancing}
\label{sec:cqb}
\begin{figure}
    \centering
    \includegraphics[width=0.8\linewidth]{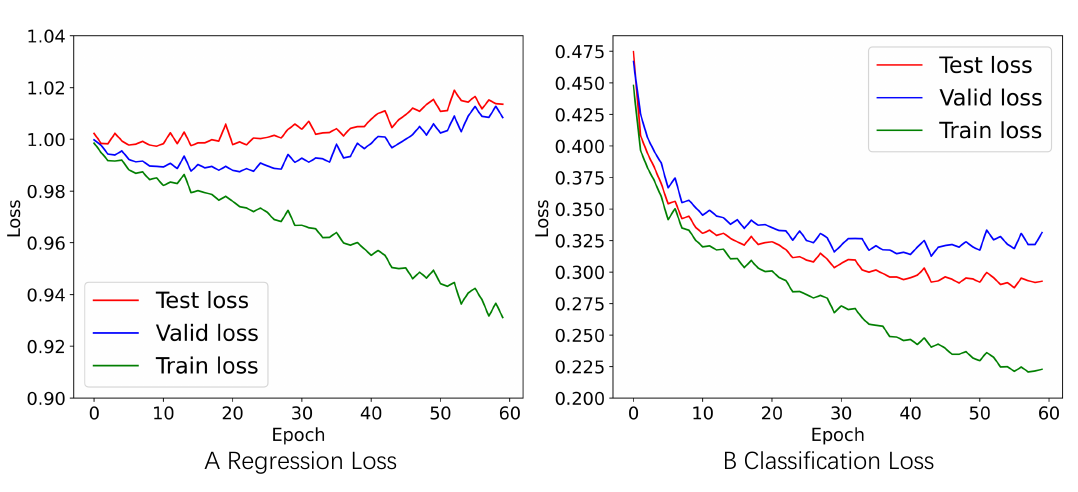}
    \caption{Training and validation losses of regression and classification tasks under single-task learning, showing clear overfitting.}
    \label{fig:overfit}
\end{figure}

As discussed in Section~\ref{sec:intro}, stock forecasting models are prone to overfitting and exhibit significant imbalance between task losses and gradient magnitudes.
Figure~\ref{fig:overfit}.B further shows that, under multi-task training, different tasks may overfit at different rates within the same epoch.
These issues motivate the need for an optimization strategy that jointly balances gradient magnitude and overfitting dynamics.
To this end, we propose Converge-based Quad-Balancing (CQB).

\textbf{EMA Balancing and Gradient Normalization.}
We first smooth task-specific gradients using an exponential moving average for gradient from multiple task separately, take regression task as an example:
\begin{equation}
    \hat{\mathbf{g}}_{r, \ell} = \beta\hat{\textbf{g}}_{r, \ell-1}+(1-\beta)\mathbf{g}_{r, \ell},
\label{eq:EMA}
\end{equation}
where $\beta \in (0,1)$ is the forgetting rate and $\mathbf{g}_{r, \ell}$ represents the gradient.
For the regression and classification tasks (momentum prediction and score forecasting), gradients are independently smoothed and then normalized to the same $l_2$ norm.
To balance the influence of regression and classification gradients while preserving their relative directions, we first normalize each task-specific gradient:
\begin{equation}
    \mathbf{u}_{r,\ell} = \frac{\hat{\mathbf{g}}_{r,\ell}}{\|\hat{\mathbf{g}}_{r,\ell}\|_2},
\quad
\mathbf{u}_{c,\ell} = \frac{\hat{\mathbf{g}}_{c,\ell}}{\|\hat{\mathbf{g}}_{c,\ell}\|_2},
\end{equation}
where the $\hat{\textbf{g}}_{r,\ell}$, $\hat{\textbf{g}}_{c,\ell}$ are gradients from regression and classification tasks after EMA smoothing in iteration $\ell$ and then rescale their combination to a comparable magnitude:
\begin{equation}
    \widetilde{\mathbf{g}}_\ell
=
\alpha_\ell \, (\mathbf{u}_{r,\ell} + \mathbf{u}_{c,\ell}),
\quad
\alpha_\ell = \max(\|\hat{\mathbf{g}}_{r,\ell}\|_2, \|\hat{\mathbf{g}}_{c,\ell}\|_2).
\label{eq:gradient_norm}
\end{equation}
This formulation ensures that neither task dominates the update due to scale differences, while keeping the overall update magnitude aligned with the larger gradient.

\textbf{Adaptive Forgetting Rate balancing.}
To mitigate overfitting during training, we introduce an adaptive mechanism that modulates the forgetting rate $\beta$ in Eq.~\ref{eq:EMA} based on observable training dynamics.
Specifically, we track the relative convergence trends between training loss and validation loss over a short temporal window, which serves as a coarse diagnostic signal indicating potential overfitting.
Importantly, the validation loss is never used for gradient computation or parameter updates. 
Instead, it is treated purely as an indicator, without introducing any validation-dependent supervision into the optimization process.

Specifically, at epoch $n$, we define a relative convergence ratio as
\begin{equation}
    V_n = {\Delta \mathcal{L}_{valid}}/{\Delta \mathcal{L}_{train}},
\label{eq:rate_eq}
\end{equation}
which could indicate the change of losses on the validation set relative to the loss on the training set.

To obtain a more stable estimate of loss variation, the loss change $\Delta \mathcal{L}$ is computed using an averaged window over several previous epochs:
\begin{equation}
\begin{split}
    \Delta \mathcal{L} = \mathcal{L}_{n-1} - mean([\mathcal{L}_{n-2b},,\mathcal{L}_{n-b-1}]) ,
\end{split}
\end{equation}
where b controls the size of the averaging window.
In practice, we set $b=6$ based on the empirical loss evolution shown Figure~\ref{fig:overfit}.
The ratio $V_n$ is computed once sufficient history is available (after 12 epochs); prior to that, $V_n$ is initialized to 1 to avoid unstable estimates.

After computing the relative converge rate for both tasks, we compute $\beta_n$ to replace the fixed $\beta$ at epoch $n$ using the following equation:

\begin{equation}
    \beta_{n} = \beta^{\sigma(V_n)}
\label{eq:beta_n}
\end{equation}
, where $\sigma(V_n)$ is a sigmoid function $\frac{1}{(1+e^{-V_n})}$.
The formulation of Eq.~\ref{eq:beta_n} follows a simple design rationale.
The forgetting rate $\beta_n$ is required to lie in (0,1) for stable EMA updates, and its adjustment should be monotonic with respect to the relative convergence signal $V_n$.
We therefore adopt a sigmoid function, which provides a smooth, bounded, and monotonic mapping and is commonly used for adaptive control in optimization.
To analyze whether our method could help delay overfitting during training, we make the following assumptions:

Assumption 1: $\Delta \mathcal{L}_{train} < 0$ during training.

Assumption 2: When overfitting occurs, the validation loss exhibits a non-decreasing trend relative to the training loss over a short temporal window.

Based on these assumptions, we make the following observation: 
Under Assumptions 1 and 2, when observable overfitting signals emerge, the adaptive forgetting rate $\beta_n$ tends to increase, which attenuates the relative contribution of current-epoch gradients to the parameter update.

In our setting, Assumption 1 is generally satisfied as the model continues to converge on the training data.
Assumption 2 reflects a commonly observed training behavior, where degradation in generalization is manifested through a divergence between training and validation losses.

\textit{Mechanistic Analysis.} 
We offer a mechanistic explanation of how CQB reacts to observable overfitting signals.
First, we analyze the change of $\beta_n$.
We define the adaptive forgetting rate $\beta_n$ as an exponential function of the convergence indicator $V_n$ in Eq.~\ref{eq:beta_n}.
Denote $s = \sigma(V_n)$, then:
\begin{equation}
    \beta_{n} = \beta^s = exp(s\ln\beta)
\label{eq:proof1}
\end{equation}
The derivative of $\beta_n$ with respect to $V_n$ is:
\begin{equation}
\begin{aligned}
        \frac{\mathrm{d}\beta_n}{\mathrm{d}V_n} &= exp(s\ln\beta)(\ln\beta)\frac{\mathrm{d}s}{\mathrm{d}V_n} \\ &=\beta^{\sigma(V_n)}(\ln\beta)\sigma(V_n)(1-\sigma(V_n)),
\end{aligned}
\label{eq:proof2}
\end{equation}
since $\sigma(V_n)(1-\sigma(V_n))>0$ for all $V_n$, then we have $sign(\frac{\mathrm{d}\beta_n}{\mathrm{d}V_n}) = sign(\ln\beta)$. Therefore when $0<\beta<1$, we have $\ln\beta<0$ and consequently $\frac{\mathrm{d}\beta_n}{\mathrm{d}V_n} < 0$, 
indicating that $\beta_n$ is monotonically decreasing with respect to $V_n$.
When overfitting occurs, $\Delta \mathcal{L}_{{train}}$ becomes negative, while $\Delta \mathcal{L}_{{valid}}$ gradually increases and may even turn positive. As a result, $V_n$ shifts from positive to negative, which in turn leads to an increase in $\beta_n$.
This property ensures that the forgetting rate reacts deterministically to overfitting signals, 
rather than remaining fixed throughout training.


We further analyze the relative weight change of gradient when overfitting happens.
We split the gradient after epoch M into two parts: gradients before and after iteration $n$ to better observe the change of gradient in the current epoch:
\begin{equation}
    \hat{\mathbf{g}}_{n+m} = \beta_n^{m+1} \hat{\mathbf{g}}_{n-1} + (1-\beta_n)\sum_{i=0}^m \beta_n^i \textbf{g}_{n+m-i}
\label{eq:EMA_expand}
\end{equation}
For the second part gradient, since $\beta \in (0, 1)$ in Eq.~\ref{eq:beta_n}, we can derive the following inequality:
\begin{equation}
      (1-\beta_n) \beta_n^i \leq (1-\beta_n), \forall i\in (0,m)
\label{eq:unqual}
\end{equation}
We can construct the ratio of $\hat{\textbf{g}}_{n-1}$ weight to $\textbf{g}_{n+m-i}$ weight for all $i$ and derive the lower bound of the ratio:
\begin{equation}
    \frac{\beta_n^{m+1}}{(1-\beta_n)\beta_n^i} \geq \frac{\beta_n^{m+1}}{1-\beta_n}, \forall i \in (0,m)
\label{eq:weight_ratio}
\end{equation}
where $\frac{\beta_n^{m+1}}{1-\beta_n}$ is the lower bound of the ratio of stable gradient
$\hat{\textbf{g}}_{n-1}$ to $\textbf{g}_{n+m-i}$ for all iterations in the current epoch.
We can easily infer that the lower bound will increase when $\beta_n$ increases while $\beta_n \in (0,1)$, as a result the ratio of gradients weight $\hat{\textbf{g}}_{n-1}$ increases.
While overfitting happens, the weight of the gradient $\beta_n^{m+1}$ before iteration $n$ contributes more to $\hat{\textbf{g}}_{n+m}$ in Eq.~\ref{eq:EMA_expand}.
This mechanism reduces the immediate influence of gradients computed after overfitting signals emerge,
thereby stabilizing optimization under non-stationary training dynamics.

\textbf{$L_2$ Regularization Balancing.} To further alleviate overfitting, CQB adaptively adjusts weight decay in the optimizer:
\begin{equation}
    decay_n = decay \cdot \sigma(-\mathrm{mean}(V_{n-1})),
\end{equation}
where $decay$ is the initial weight decay.
As validation performance deteriorates, stronger regularization is applied to stabilize training.

%% file: 04experiment.tex

\begin{table*}[t]
\caption{Experimental results on SEE50, CSI100, and CSI300. The best and second-best results are highlighted in \textbf{bold} and \underline{underlined}. Every group is repeated three times and the standard deviation ($\times10^{-3}$) is reported.}
\centering
\resizebox{0.7\textwidth}{!}{%
\begin{tabular}{clrrrrrrrr}
\toprule
\multirow{2}{*}{\textbf{Backbone}} & \multirow{2}{*}{\textbf{Method}} & \multicolumn{2}{c}{\textbf{SEE 50}}  && \multicolumn{2}{c}{\textbf{CSI 100}}                                   &  & \multicolumn{2}{c}{\textbf{CSI 300}}          \\ \cline{3-4} \cline{6-7} \cline{9-10} 
& & IC$\uparrow$     & RankIC$\uparrow$ && IC$\uparrow$     & RankIC$\uparrow$    && IC$\uparrow$   & RankIC$\uparrow$    \\ \hline

\multirow{5}{*}{LSTM}      & STL  & \underline{0.0272(0.6)}& 0.0276(1.3)&   & \underline{0.0493(2.9)}&\underline{0.0438(2.6)}         &     & \underline{0.0620(2.0)}  & \underline{0.0586(1.7)}   \\
& EW  &0.0270(0.6)&0.0271(1.1)&       & 0.0490(1.9)  & 0.0431(2.8)       &  & 0.0571(1.2) & 0.0538(1.5)    \\
& DB-MTL & \underline{0.0272(1.7)}&0.0279(0.8)& & 0.0470(3.4) & 0.0415(3.0)& & 0.0567(3.0)& 0.0543(2.9)    \\
& CAGrad  &0.0267(2.0)& \underline{0.0300(3.1)}&  & 0.0469(1.4) & 0.0422(0.5)  &  & 0.0551(0.9)& 0.0528(1.0)\\
& MiM-StocR (ours)   & \textbf{0.0362(3.1)}& \textbf{0.0358(3.6)} &  & \textbf{0.0522(1.6)} & \textbf{0.0467(1.5)}& & \textbf{0.0632(0.9)}& \textbf{0.0604(1.4)} \\ \hline

\multirow{5}{*}{GATs}      & STL &0.0258(3.4)&0.0261(1.3)&  & 0.0421(6.3)& 0.0360(5.7) &     & 0.0575(1.5)    & 0.0546(1.5)   \\
& EW    &\underline{0.0269(3.8)}&\textbf{0.0280(4.9)}&     & 0.0386(6.8) & 0.0339(5.4)   &    & 0.0588(2.6) & 0.0556(2.4)    \\
& DB-MTL &0.0242(2.6)&0.0219(4.0)&  & 0.0394(5.7) & 0.0358(5.3)  &  & 0.0589(3.2)& 0.0566(2.7) \\
& CAGrad  &0.0223(3.1)&0.0182(5.0)&  & \underline{0.0423(3.5)} & \underline{0.0378(2.6)}& & 0.0608(1.8) &0.0588(1.2)\\
& MiM-StocR (ours)       &\textbf{0.0278(4.8)}& \underline{0.0266(4.4)}&                     & \textbf{0.0472(8.7)} & \textbf{0.0443(6.9)} &     & \textbf{0.0622(1.5)}& \textbf{0.0590(0.8)}   \\ \hline

\multirow{5}{*}{HIST}     & STL        &0.0288(0.8)&\underline{0.0300(1.7)}&              & 0.0552(1.8) & 0.0503(1.2)&     & \textbf{0.0672(2.3)} & \underline{0.0630(2.3)}      \\
& EW   &0.0286(0.5)&0.0297(1.2)&      & \underline{0.0571(2.4)} & 0.0512(2.2)   &     & 0.0631(1.5) & 0.0601(1.4)   \\
& DB-MTL          &0.0278(3.8)&0.0289(1.9)&             & 0.0565(1.4) & \underline{0.0517(1.9)}      &  & 0.0631(1.8)& 0.0599(1.9)    \\
& CAGrad      &\underline{0.0301(2.4)}&0.0292(0.6)&              & 0.0560(1.3)   & 0.0507(1.4)      &  & 0.0638(3.4) & 0.0611(2.7)     \\
& MiM-StocR (ours)      &\textbf{0.0393(2.3)}  & \textbf{0.0387(3.6)}&                 & \textbf{0.0605(1.1)} & \textbf{0.0544(2.0)}    &     & \underline{0.0667(1.1)}& \textbf{0.0633(1.0)}  \\ \bottomrule
\end{tabular}
}
\label{tab:experiment_result}
\end{table*}

In this section, we evaluate MiM-StocR with extensive experiments.
We start with an introduction of baselines, implementations and datasets. 
Then we discuss the performance of MiM-StocR and the effectiveness of each components.

\subsection{Experimental Setting}

\textbf{Backbones and Baselines.} We evaluate on three representative backbones: LSTM~\cite{hochreiter1997long}, GATs~\cite{velickovic2017graph}, and HIST~\cite{xu2021hist}. 
HIST is one of the best performing stock recommendation model in Qlib.
We compare with standard training strategies and state-of-the-art multi-task optimization methods, including Single Task Learning (STL), Equal Weighting (EW), DB-MTL~\cite{lin2023dual}, and CAGrad~\cite{liu2021conflict}.

\textbf{Implementation Details.} We set the momentum-line parameters to $l=4$ and $s=6$ in momentum label construction as the supervision signals. For both classification and regression tasks, we adopt MSE for regression and the objective in Section~\ref{sec:ndcg} for classification.
In Adaptive-$k$ ApproxNDCG, we set the minimum truncation threshold to $20\%$ of the stock pool size, consistent with common portfolio construction practices.
In CQB we use an initial weight decay of $10^{-3}$ and a forgetting rate $\beta=0.5$ to start the training.
Models are trained with a learning rate of $2\times10^{-4}$ for up to 100 epochs until best validation loss stops decreasing. For statistical significance, each experiment is conducted three times under identical settings.
All backbones have fewer than 500M parameters and can be trained on a RTX 3090 GPU. The average training time per baseline is approximately 3 hours.

\textbf{Stock Price Dataset.} Following~\cite{xu2021hist}, we conduct experiment on the Alpha360 data, including SEE50, CSI100, and CSI300.
All datasets are preprocessed by normalization and split into training (2007--2014), validation (2015--2016), and test (2017--2020) sets.
This ensures no test data leakage or look-ahead bias.
We adopt Information Coefficient (IC) and RankIC~\cite{xu2021hist,lin2021learning} as evaluation metrics.
IC measures the Pearson correlation between predicted scores and one-day returns, while RankIC computes the same correlation on ranked returns.
Both metrics range from $[-1,1]$, with higher values indicating better forecasting and ranking performance.
To assess profitability, we backtest on CSI300 using Qlib’s default Top50 strategy, which buys the top-ranked stocks daily and sells them on the next trading day~\cite{li2019multi}.

\subsection{Main Experimental Results}

\begin{figure}[t]
\centering
\includegraphics[width=0.85\linewidth]{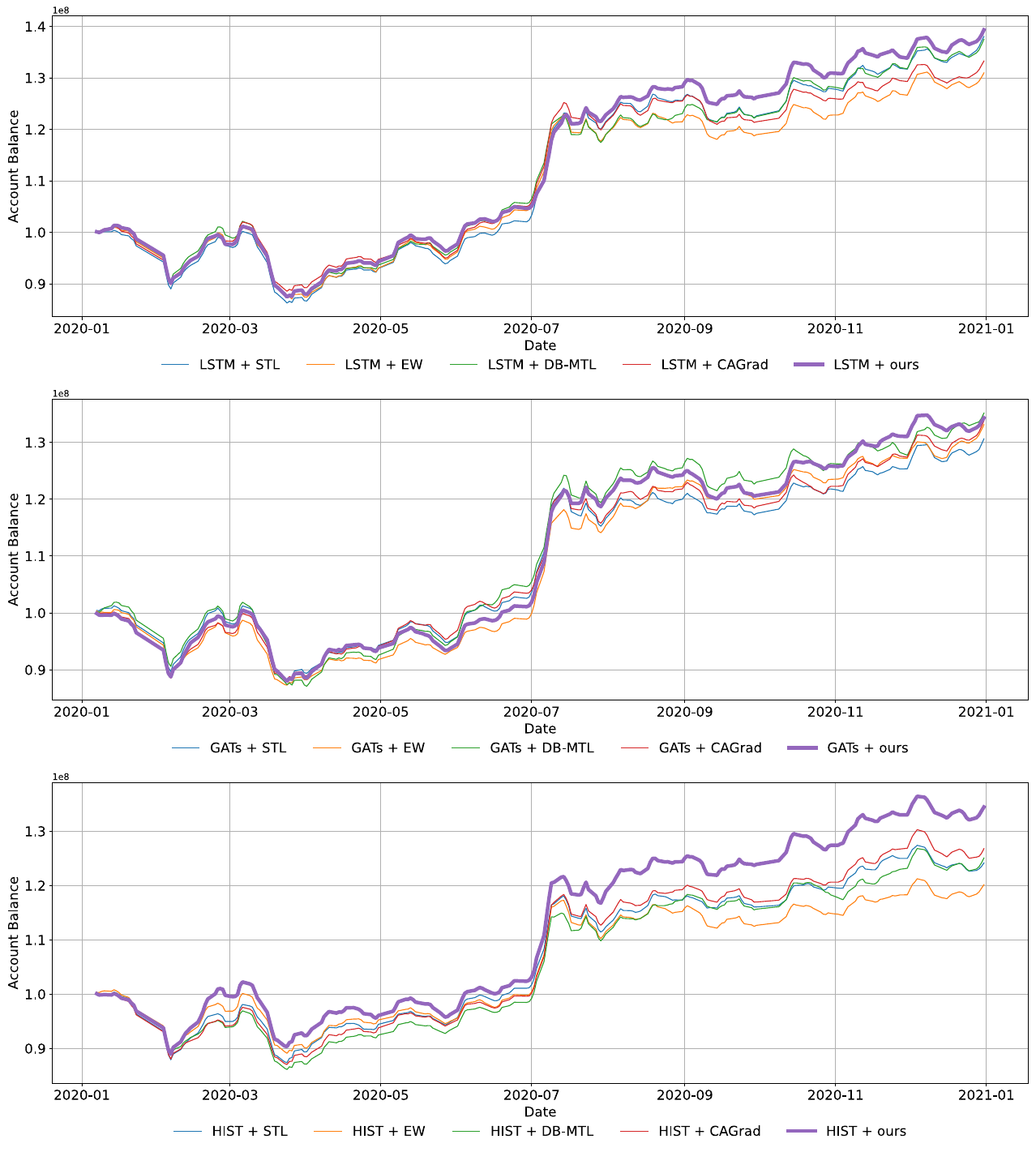}
\caption{Cumulative returns on CSI300. MiM-StocR achieves the highest profit, outperforming the index and multi-task baselines.}
\vspace{-0.2cm}
\label{fig:backtest}
\end{figure}

We conduct experiments on three backbones and datasets using proposed methods and baselines. The experimental results are shown in Table~\ref{tab:experiment_result}.
In all nine combinations of datasets and backbones, \textbf{MiM-StocR outperforms all baselines}, which indicates that our framework could enhance the backbone's performance in forecasting the return ratio and ranking. Our method outperforms all baselines on stock data of different scales, demonstrating the strong generalization and robustness of the proposed method.
Regarding the profit evaluation, we plot the change in account balance during the simulation in Figure~\ref{fig:backtest}. 
The combination of LSTM and ours method achieves the highest profit, which is 11.6\% higher than the CSI300 index. In trading simulations with different backbones, our method consistently achieves top investment returns, demonstrating its robustness in profitability.

\begin{table}[t]
\centering
\caption{Comparison of momentum-based and rise-or-fall classification under identical settings.}
\vspace{-0.2cm}
\resizebox{0.7\linewidth}{!}{
\begin{tabular}{llll}
\hline
Backbone              & Task         & IC $\uparrow$     & RankIC $\uparrow$         \\ \hline
\multirow{2}{*}{LSTM} & Rise-or-Fall & 0.0457(4.5) & 0.0436(4.5)  \\
                      & Momentum      & \textbf{0.0632(0.9)}& \textbf{0.0604(1.4)} \\ \hline
\multirow{2}{*}{GATs} & Rise-or-Fall  & 0.0501(3.4)& 0.0484(3.9) \\
                      & Momentum     & \textbf{0.0622(1.5)}  & \textbf{0.0590(0.8)}\\ \hline
\multirow{2}{*}{HIST} & Rise-or-Fall  & 0.0519(2.6) & 0.0507(2.2)\\
                      & Momentum     & \textbf{0.0667(1.1)}  & \textbf{0.0633(1.0)}\\ \hline
\end{tabular}
}
\label{tab:rise-fall-comp}
\end{table}

\subsection{Ablation Study}

\textbf{Impact of Momentum Line.}
\label{sec:rq2}
To validate the effectiveness of the momentum signal,  
we conduct experiments using rise-or-fall tasks~\cite{chong2017deep} to replace the momentum line task in MiM-StocR. The experimental results using the rise-or-fall classification are shown in Table~\ref{tab:rise-fall-comp}. 
Under the same experimental settings, the rise-or-fall task achieves lower IC and RankIC, which indicates that training with rise-or-fall tasks fails to improve the models' perception of price directions.
The deterioration of performance could be attributed to the noisy nature of rise-or-fall signal that has been noticed by researchers for decades~\cite{bengio1997using}, and the accurate prediction of rise-or-fall is extremely challenging~\cite{deng2019knowledge,hu2018listening}. 

\begin{table}[t]
\centering
\caption{Precision@N of different classification objectives. Adaptive-k ApproxNDCG achieves the best performance.}
\resizebox{0.7\linewidth}{!}{
\begin{tabular}{lccccccccc}
\toprule
Precision & @10 & @20 & @30 & @50 \\ \hline
Cross-entropy         & 53.58 & 53.98 & 53.99 & 53.67 \\
Pair-wise & 54.07 & 54.15 & \textbf{54.15} & 53.82 \\
w/o adaptive-k         & 54.04 & 54.01 & 53.93 & 53.64 \\
MiM-StocR & \textbf{54.42}  & \textbf{54.33}  & \textbf{54.15}  & \textbf{53.84}  \\ \bottomrule
\end{tabular}
}
\label{tab:ndcg_compare}
\end{table}

\textbf{Effect of Adaptive-k ApproxNDCG.}
A series of ablation experiments were conducted to illustrate the effectiveness of the Adaptive-k ApproxNDCG objective function:
\textbf{Pair-wise}: NDCG part is replaced by pair-wise loss function~\cite{feng2019temporal};
\textbf{Cross-entropy}: NDCG part is removed, and only cross-entropy is used as the loss function; 
\textbf{w/o Adaptive-k}: Adaptive-k mechanism is removed and $k$ is fixed to 50, the exact number of Top50 trading strategy. 

The results are shown in Table~\ref{tab:ablation experiment}, all groups have been outperformed by our method.
The results indicate that proposed objective function could introduce rank-related information to improve performance. 
To investigate whether applying Adaptive-k ApproxNDCG can enhance the model's perception of stock rankings, especially top stocks, we compute Precision@N for these methods.
The precision@N is the proportion of top N stocks ranked by predicted scores, and the one-day return ratio is positive (rise). We set N to 10, 20, 30, and 50 and compute the Precision@N of different objective functions in Table~\ref{tab:ndcg_compare}.
The results indicate that our method could improve the ratio of profitable stocks at the top of the recommendation.
Using the Adaptive-k ApproxNDCG has improved the backbones' ability to identify the rank and direction of stocks.


\begin{table}
\centering
\caption{Ablation study on CSI300 (HIST backbone), evaluating ranking objectives and CQB components.}
\resizebox{0.7\linewidth}{!}{%
\begin{tabular}{lrr}
\hline
& IC $\uparrow$     & RankIC  $\uparrow$   \\ \hline
\multicolumn{3}{l}{\textbf{RQ3: Diff. objective functions}}\\
Cross-entropy  & 0.0640(3.0)& 0.0612(3.0)    \\
Pair-wise  & 0.0657(1.7) & \underline{0.0625(2.2)} \\
w/o Adaptive-k   & 0.0649(0.2) & 0.0618(0.5)     \\ \hline
\multicolumn{3}{l}{\textbf{RQ4: Diff. multi-objective optimizations}} \\
w/o $\beta$ balancing   & 0.0656(1.0)  & 0.0619(1.1)    \\
\vspace{0.03cm}
w/o L2 balancing    & \underline{0.0665(2.3)}   & \underline{0.0625(2.4)} \\ \hline
MiM-StocR     & \textbf{0.0667(1.1)} & \textbf{0.0633(1.0)} \\ \hline
\end{tabular}
}
\label{tab:ablation experiment}
\end{table}

\begin{figure}[t]
\centering
\includegraphics[width=0.8\linewidth]{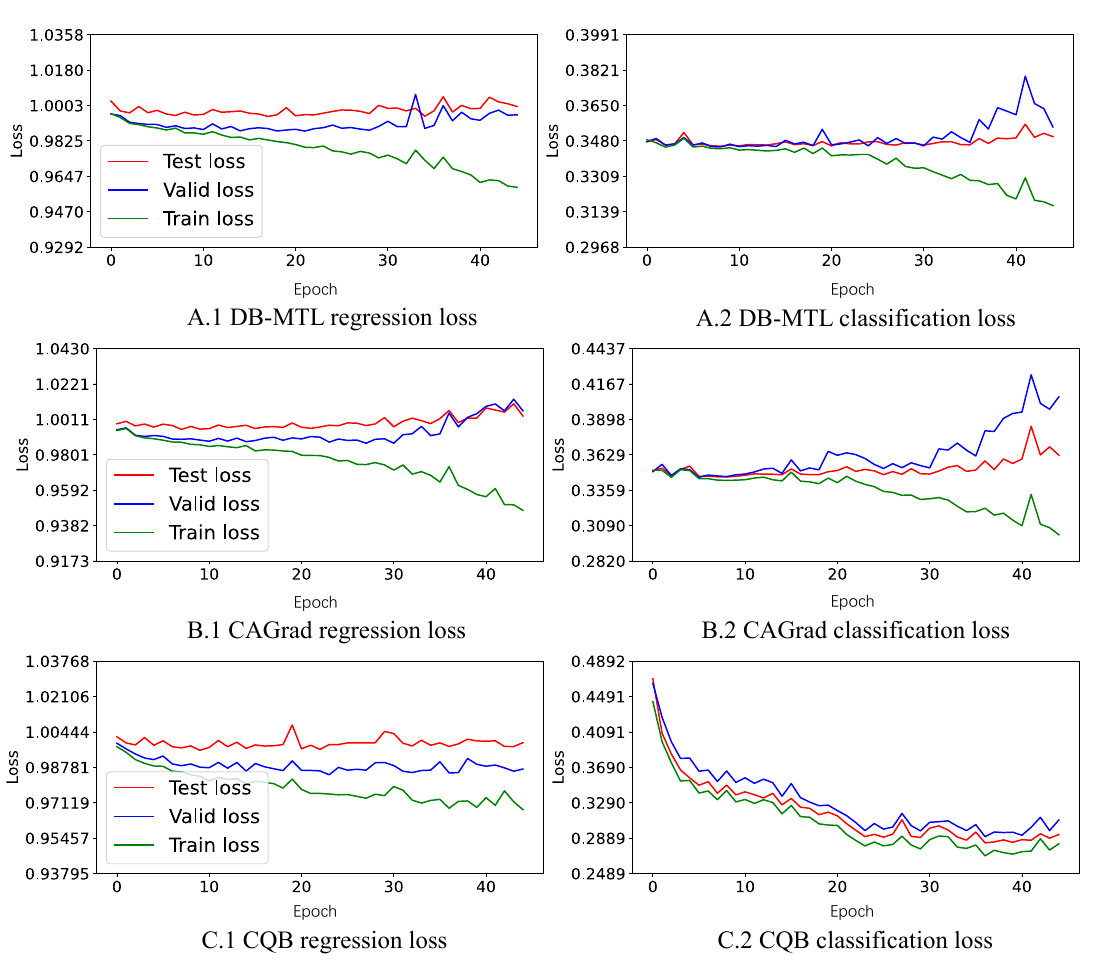}
\caption{Train and test losses under different multi-task optimizers. CQB delay the overfitting compared with existing methods.}
\vspace{-0.4cm}
\label{fig:cqb_loss}
\end{figure}

\textbf{Converge-based Balancing under Overfitting Dynamics.} We analyze the effect of CQB on training dynamics by visualizing task-specific losses and conducting ablation studies. 
Figure~\ref{fig:cqb_loss} compares the training and test losses of CQB with baselines.
Across baselines, we observe a common pattern in which test losses start to increase after approximately 25 epochs, while training losses continue to decrease, reflecting overfitting.
With CQB, the divergence is delayed and its magnitude is reduced compared to existing methods.
This indicates that CQB attenuates the influence of overfitted gradients, which is expected in non-stationary financial time series.
In addition to visual comparisons, we conduct ablation studies to assess the contribution of individual components in CQB (Table~\ref{tab:ablation experiment}).
In \textbf{w/o $\beta$ balancing}, a uniform $\beta$ is used throughout training, while in \textbf{w/o L2 balancing}, the weight decay is fixed.
Disabling either component consistently degrades performance, suggesting that adaptive forgetting-rate control and regularization balancing jointly contribute to more stable training across tasks.

\subsection{Hyperparameter Sensitivity Analysis}

\begin{figure}[t]
\centering
\includegraphics[width=\linewidth]{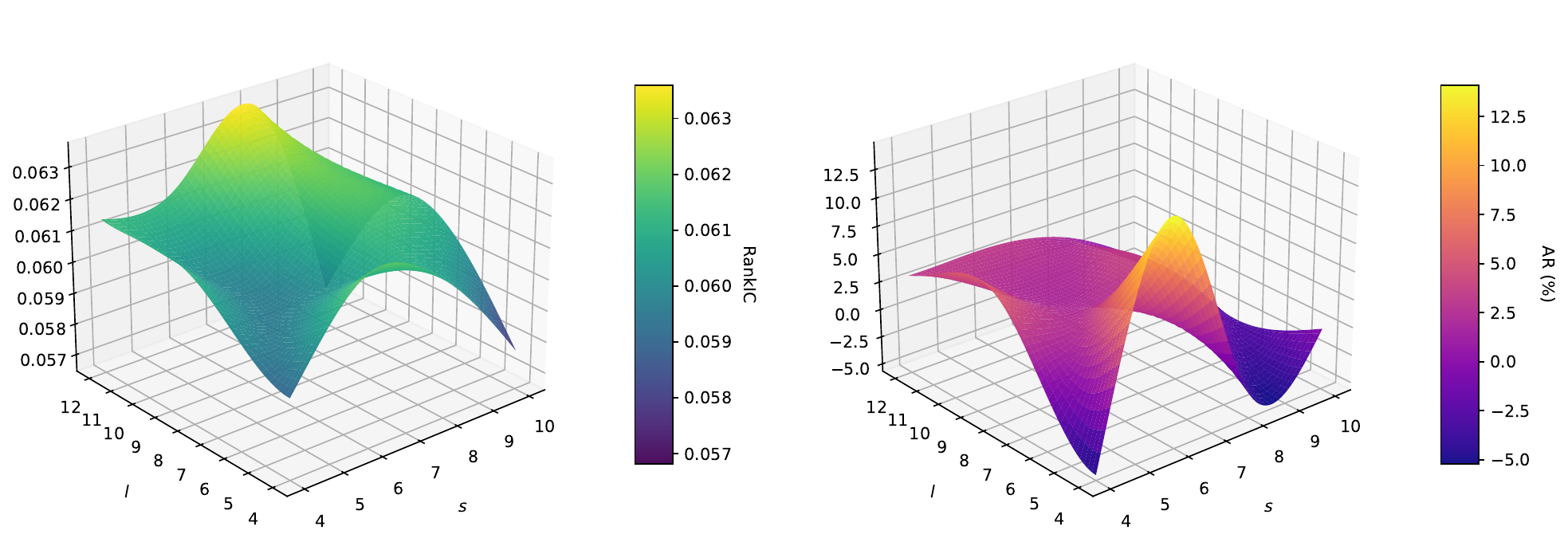}
\caption{RankIC performance under different momentum-line parameters on CSI300. Short-term settings (e.g., $s=6, l=4$) yield strong performance.}
\vspace{-0.4cm}
\label{fig:hyperparameter1}
\end{figure}

\begin{figure}[t]
\centering
\includegraphics[width=0.6\linewidth]{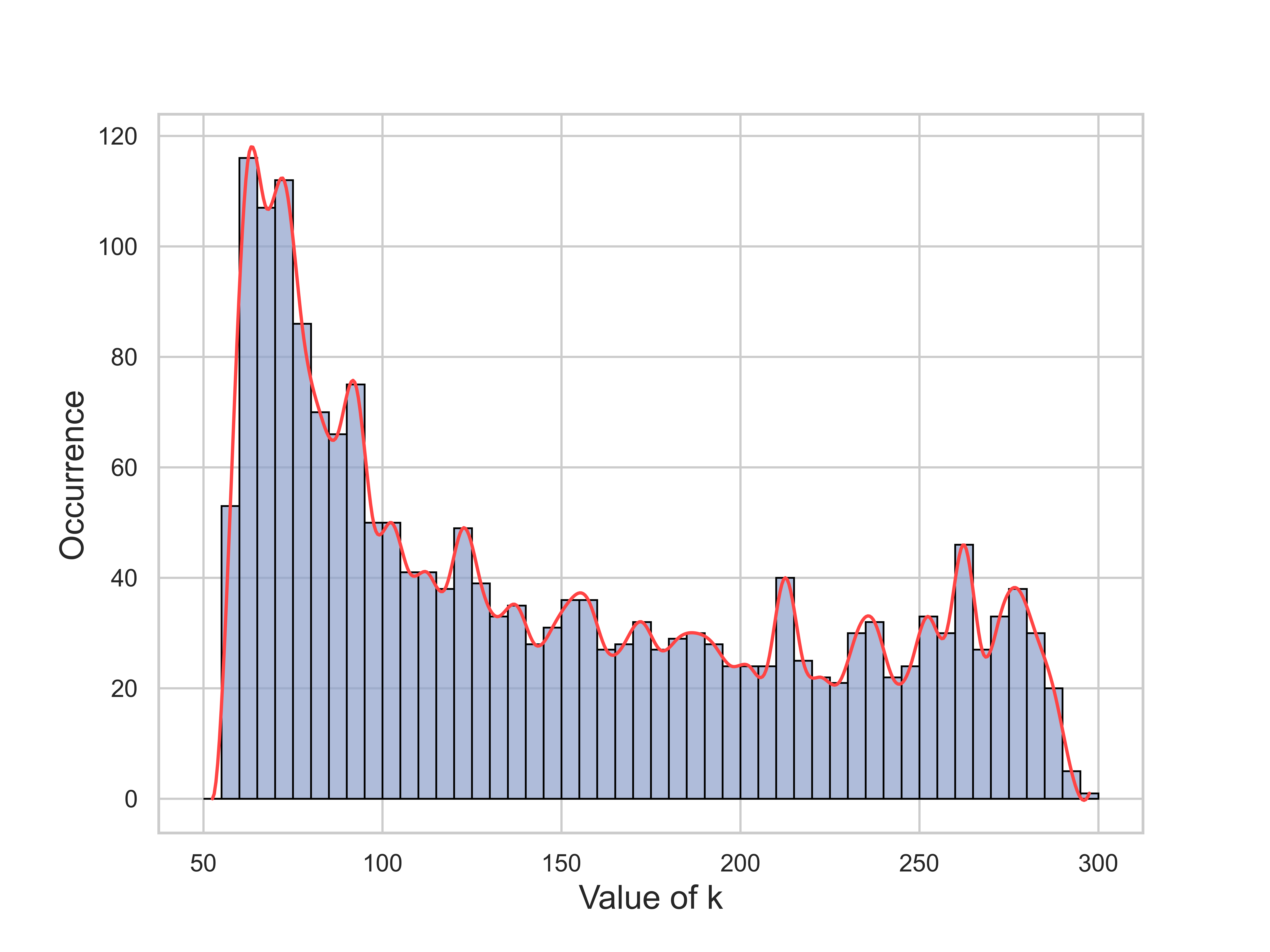}
\caption{Distribution of adaptive k values across trading days, showing dynamic adjustment to market conditions.}
\vspace{-0.4cm}
\label{fig:hyperparameter2}
\end{figure}

\textbf{Momentum Line Parameters.}
The momentum line is controlled by two parameters: the interval $l$ between price points and the line length $s$, which together determine the temporal scope of short-term trend modeling.
We consider $l \in \{4,8,12\}$ and $s \in \{4,6,8,10\}$ to cover typical weekly to multi-week horizons while avoiding overly short or excessively diluted trends.
Experiments on CSI300 with the HIST backbone in Figure~\ref{fig:hyperparameter1} show that larger $(s,l)$ combinations can achieve slightly higher RankIC, while $(s=6,l=4)$ yields the best performance in profit-oriented backtesting.
Balancing ranking accuracy and profitability, we adopt $s=6$ and $l=4$ as the default configuration in all experiments.

\textbf{Visualization of Adaptive-$k$ during Training.}
Figure~\ref{fig:hyperparameter2} illustrates the distribution of adaptive $k$ values during training.
The value of $k$ varies with the daily momentum distribution, increasing when a large proportion of stocks share the same high-momentum category.
This prevents truncation bias by ensuring that stocks with identical momentum levels are not split by a fixed cutoff.
We set the minimum $k$ threshold to $20\%$ of the stock pool, consistent with common investment practices on CSI300.

%% file: 06conclusion.tex
In this work, we propose MiM-StocR, which integrates a momentum-line auxiliary task, an Adaptive-k ApproxNDCG objective, and CQB optimization to improve generalization and mitigate overfitting.
We incorporate the momentum line as an auxiliary signal to provide more structured and less noisy trend information than simple rise-or-fall labels.
To better align learning objectives with ranking-oriented investment decisions, we adopt an adaptive top-k variant of Approx-NDCG to emphasize the importance of highly ranked stocks.
In addition, we explore a gradient balancing strategy based on historical gradients to improve training stability and delay overfitting in multi-task settings.
Future research may further investigate how to exploit diverse auxiliary tasks and integrate heterogeneous outputs within multi-task learning to achieve improved robustness and risk-aware performance.